\documentclass[preprint, 10pt]{aastex}
\shorttitle{NGC~604}
\shortauthors{Miskey et al.}

\begin{document}

\title{STIS Spectral Imagery of the OB Stars in NGC~604: Describing the
Extraction Technique for a Crowded Stellar Field}

\author{Cherie L. Miskey\altaffilmark{1} and Fred C. Bruhweiler\altaffilmark{1}}
\affil{Institute for Astrophysics \& Computational Sciences}
\affil{Department of Physics, The Catholic University of America, Washington, DC 20064}
\email{miskey@iacs.gsfc.nasa.gov, bruhweiler@cua.edu}

%% ...alternate affiliations, which are identified by the \altaffilmark 
%% after each name.  Specify alternate affiliation information with 
%% \altaffiltext, with one command per each affiliation.

\altaffiltext{1}{Laboratory for Astronomy and Solar Physics, 
NASA/Goddard Space Flight Center, Greenbelt, MD 20771}

\begin{abstract}

We have developed a data reduction procedure to extract multiple spectra from a single two-dimensional Space Telescope Imaging Spectrograph (STIS) image of a crowded stellar field.  This paper provides a description of our new technique, utilizing a STIS ultraviolet spectral image, acquired with the G140L grating and the 52$\arcsec \times 2\arcsec$ aperture, sampling a concentration of O and B stars in the central region of the NGC~604 starburst in M33.  The software routines can disentangle and produce reliable ultraviolet spectra of stars with angular separations as small as 0.055$\arcsec$.  Use of the extraction slit, based upon our model of the spectral cross dispersion profile, generates spectra with slightly higher resolution than the STScI standard processing.  Our results clearly show that the spectral imaging capability of STIS represents a powerful tool for studying luminous stars in the star-forming regions of the Local Group. 
\end{abstract}

%% See the instructions to authors for the journal to which you are 
%% submitting your paper to determine what keyword punctuation is appropriate.


%\notetoeditor{}

% Here are the section headings
%\section{}
%\subsection{}
%\subsubsection{}
%\paragraph{}

\section{INTRODUCTION}

One of the designed functions of the Space Telescope Imaging Spectrograph (STIS) aboard the Hubble Space Telescope (HST) is to obtain two-dimensional stellar spectroscopy similar to objective prism spectroscopy from the ground.  However, the high angular resolution capabilities of the HST and STIS are far better than can be obtained by comparable instrumentation from ground-based observatories.  Theoretically, spectra of stars with separations of less than 0.1$\arcsec$ can be disentangled and reliably studied.  The STIS has another advantage in that it can perform ultraviolet through optical (1150-10,000~\AA) spectroscopy of individual stars in star-forming regions anywhere in the Local Group.  Despite these instrumental capabilities, no significant two-dimensional studies have been made of crowded stellar fields using the STIS.  This has been in part because of the lack of any data reduction and analysis software capable of extricating the overlapping spectra of stars in a densely populated field-of-view of a STIS spectral image. 

In this paper, we describe our data reduction software developed for processing STIS spectral images of crowded stellar fields acquired with the G140L grating and the large 2$\arcsec$ wide aperture.  Specifically, we present a description of the software and techniques used in reducing and analyzing a STIS ultraviolet spectral image of the OB association, NGC~604.

We observed with STIS the UV-bright concentration of O and B stars in NGC~604, the luminous star-forming region located in the spiral arms of the nearby galaxy M33.  NGC~604 is one of the brightest H II regions in the local Universe.  Furthermore, its close proximity (d= 840 kpc) makes it an excellent target to probe the details of enhanced star formation activity.  Our objective was to observe and resolve the young, embedded stars in the luminous H II region and obtain ultraviolet spectra of large numbers of O and B stars from a single image.  A full discussion of the subsequent analysis of these data are presented in the companion paper, Bruhweiler, Miskey, \& Smith Neubig (2003; Hereafter Paper 2).

When obtaining an image with multiple objects in a single exposure, the overlapping spectra provide particular data reduction challenges.  This is certainly the case for our STIS data of NGC~604.  In what follows, we describe aspects of data acquisition that affect the data reduction and discuss the basic technique used to produce accurate flux and wavelength calibrated spectra.  We further present examples that give an assessment of the data quality of the extracted spectra.

\section{OBSERVATIONS}

\begin{figure}
%\vspace{8 in}
%\epsscale{1}
\plotone{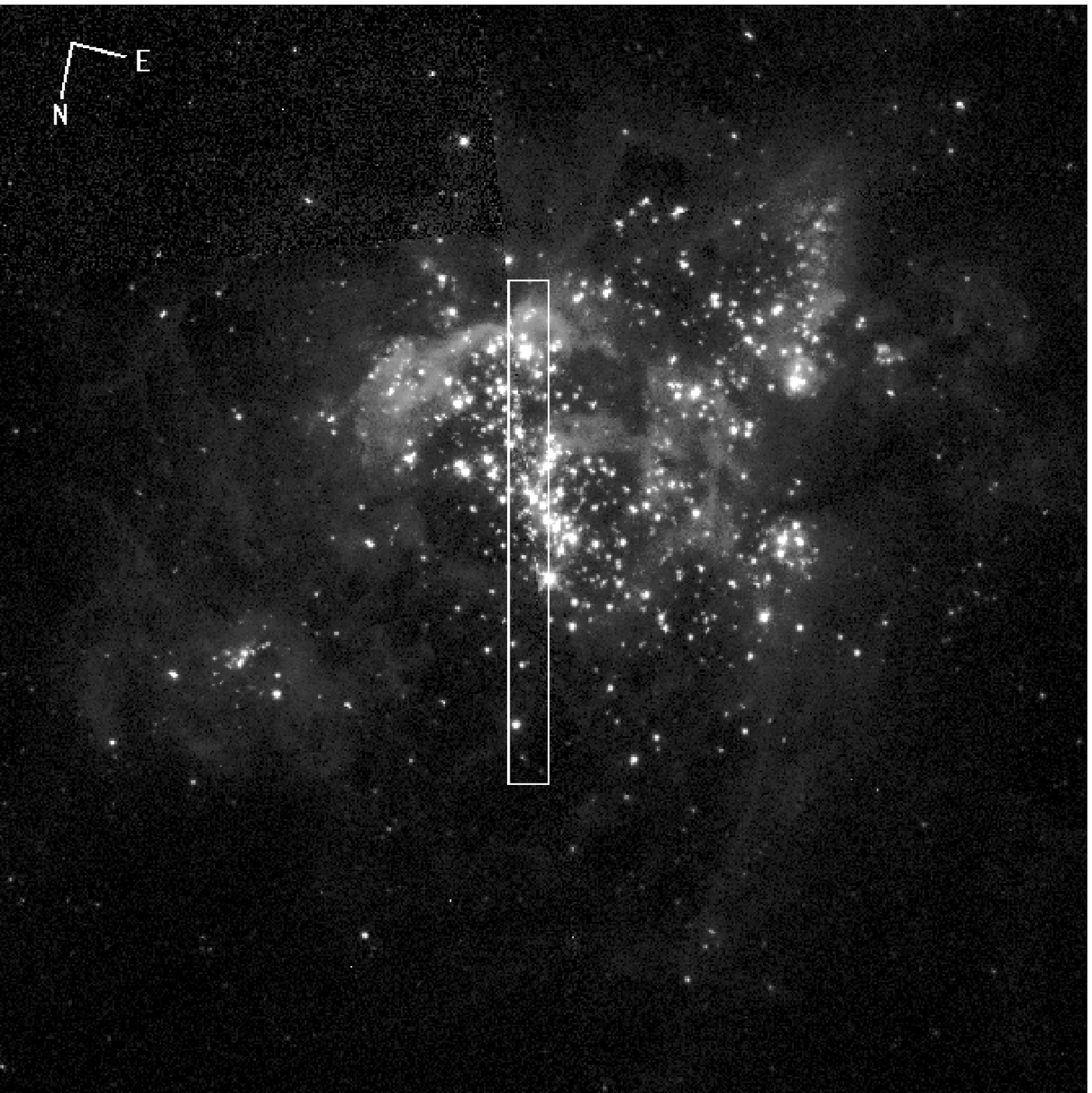}
\caption{Outline of the 2$\arcsec$ wide aperture from our STIS observation superposed upon WFPC2/F336W imagery of NGC~604.  The position angle of the aperture is -165.009 degrees East of North.  \label{fig1}}
\end{figure}

We obtained a two-dimensional, ultraviolet, spectral image (dataset ID:  o4xl01040) of the brightest concentration of stars in NGC~604 utilizing the STIS aboard HST on 1998 August 27.  A 1720 second exposure was acquired with the FUV-MAMA detector, G140L grating, and 52$\arcsec \times 2\arcsec$ aperture (that effectively samples a 25$\arcsec \times 2\arcsec$ field-of-view (FOV) for the MAMA detector).  The aperture was aligned with the position angle of the image y-axis at P.A. = -165.009 degrees East of North to maximize the number of bright stars in the slit, based on previous pre-COSTAR HST Wide Field Planetary Camera (WFPC) imagery, as presented in \citet[]{5}, and Wide Field Planetary Camera 2 (WFPC2)  imagery (see Table 1).  See  Figure \ref{fig1} for an overlay of our STIS aperture on an archival WFPC2 image taken with the F336W filter.  (Also, see Paper 2.)  For stars positioned along the long axis of the aperture, the low resolution ($\sim$ 2-3~\AA) G140L grating provides an effective wavelength range from 1150 to 1736~\AA.  The plate scale of the 1024 $\times$ 1024 pixel image is 0.0244 arcsec/pixel for the FUV-MAMA.  

Although spectral images can be obtained of a larger number of stars using the slitless mode, the geocoronal emission would contaminate the entire FOV, greatly increasing the background and lowering the signal-to-noise.  The 2$\arcsec$ wide aperture is wide enough to include a large fraction of the O and B stars in NGC~604, but still restrict the region of the spectrum affected by the high Ly$\alpha$ background.  See Section 3.4 for more discussion of the astronomical background.
 
\begin{figure}
\plotone{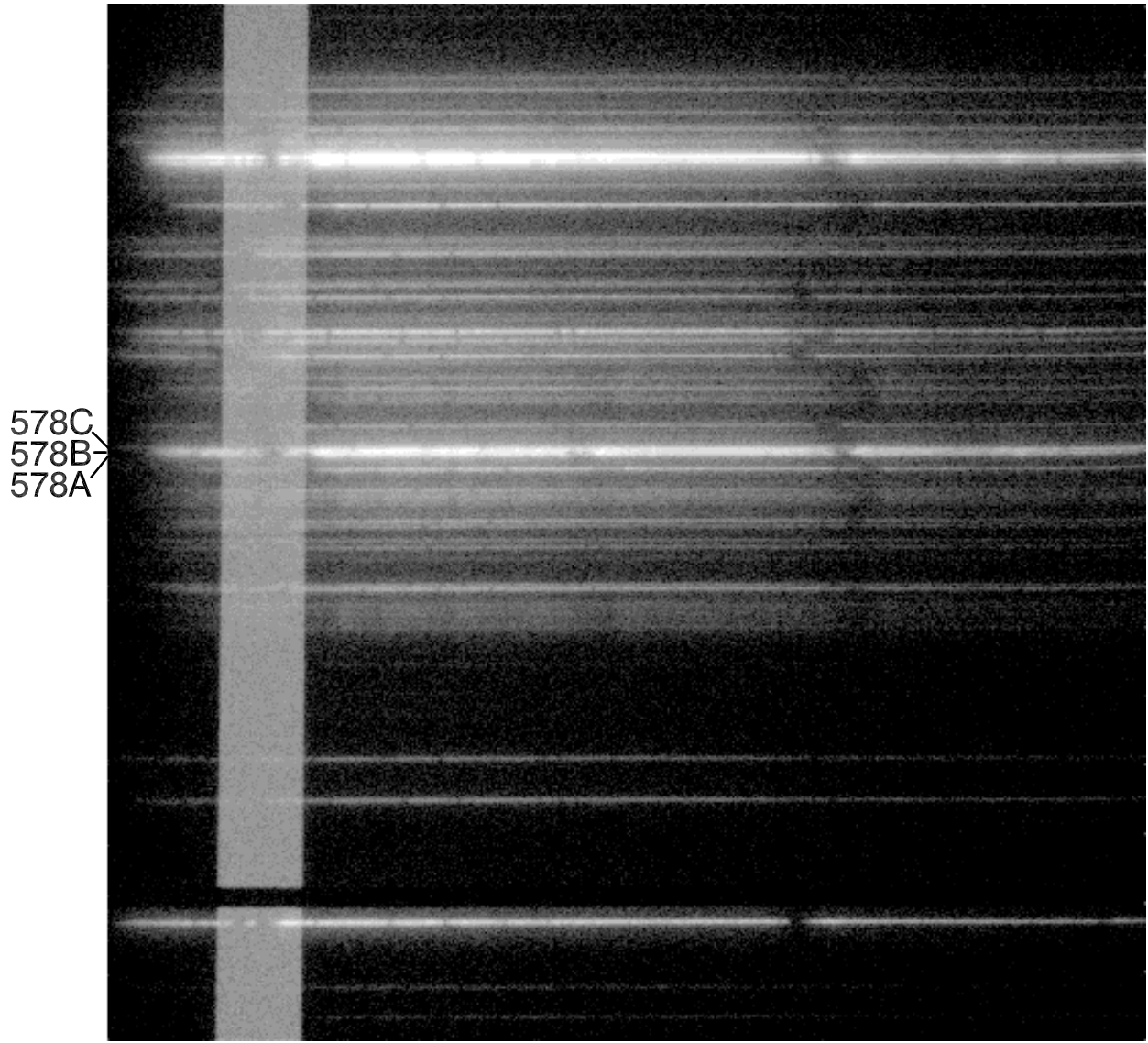}
\caption{Our STIS/G140L spectral image of bright O and B stars of NGC~604 with stellar complex 578 denoted.  The vertical band to the left in the spectral image delineates the H~I-Ly$\alpha$ geocoronal emission.  \label{fig2}}
\end{figure}

Our data show at least 60 UV spectra of O and B stars (see Figure \ref{fig2}). 
Geocoronal H~I-Ly$\alpha$ emission outlines the full 25$\arcsec$ length of the MAMA FOV of the 2$\arcsec$ wide STIS aperture.  Each spectrum clearly shows sharp interstellar absorption produced by gas in both the Milky Way and M33 galaxies.  These components are unresolved spectroscopically, since the radial velocities of M33 and NGC~604 are v$_{r}=-$179~km~s$^{-1}$ and $-$226~km~s$^{-1}$, respectively \citep[]{6}, and appear as a merged single velocity component at the G140L resolution.\footnotemark
\footnotetext{The radial velocities are from the NASA/IPAC Extragalactic Database (NED) which is operated by the Jet Propulsion Laboratory, California Institute of Technology, under contract with the NASA.  NED is available at:  http://nedwww.ipac.caltech.edu.}   These sharp interstellar lines include Si~II~$\lambda\lambda$1190,1193, N~I~$\lambda\lambda$1198,1199,1200, Si~III~$\lambda$1206, Si~II~$\lambda$1260, O~I~$\lambda$1302, Si~II~$\lambda$1304, C~II~$\lambda\lambda$1334,1335, Si~IV~$\lambda\lambda$1393,1402, Si~II~$\lambda$1526, C~IV~$\lambda\lambda$1548,1550, Fe~II~$\lambda$1608, and Al~II~$\lambda$1670.  The interstellar Si~IV and C~IV features are superposed upon broad stellar P~Cygni mass loss features.  Other prominent broad features include interstellar H~I-Ly$\alpha$, as well as stellar P~Cygni N~V~$\lambda\lambda$1238,1240 and He~II~$\lambda$1640 emission.  The actual x-positions on the MAMA detector of the spectral lines are shifted, corresponding to the x-positions of the stellar sources in the 2$\arcsec$ wide aperture.

Besides the HST imagery of NGC~604, we incorporated STIS spectral images of other individual stars to both design and test our data reduction software.  See Table 1 for a list of the relevant datasets.

\section{DATA REDUCTION}

\subsection{MODELING THE STIS CROSS DISPERSION PROFILE}

%Due to the many steps, subroutines, and data products involved in our data reduction process, several supplements are presented for reference.  See Figure \ref{fig3} for a simplified flowchart summarizing the cross dispersion profile modeling.  A glossary of terms (names of variables, arrays, and subroutines) is also included in Appendix A. 

\begin{figure}
%\epsscale{1.0}
\plotone{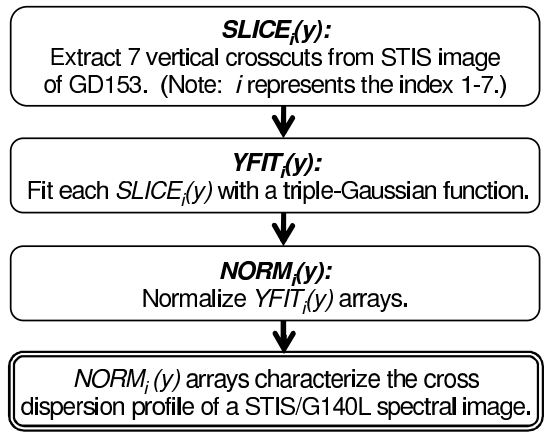}
\caption{Modeling the STIS cross dispersion profile with a MAMA/G140L spectral image of white dwarf GD153. (See Section 3.1 for further detail.) \label{fig3}}
\end{figure}

In this section, we describe the process for modeling the cross dispersion profile (CDP) of a STIS spectrum. To facilitate our discussion, in Figure \ref{fig3} we present a flowchart summarizing the steps. Because of the many subroutines and data products used in our data reduction, we also provide Appendix A, which contains a glossary of our terms. The reader should feel free to refer to Appendix A as necessary. We give a more detailed discussion of how and why we perform the CDP modeling below.

To separate closely spaced spectra on the MAMA detector face of the STIS, it is first necessary to accurately model an isolated spectrum produced by the MAMA/G140L configuration.  This will generate key parameters defining the shape and variation across the detector of the cross dispersion profile of the spectrum.  We have selected a 1024 pixel $\times$ 1024 pixel STIS/G140L spectral image of the bright, nearby, hot, white dwarf, GD153 (ID: o4a502060; See Table 1).  We choose the UV spectrum of this star to model because GD153 represents a STIS calibration target and is a continuum source with few photospheric or interstellar lines. 

All STIS/G140L datasets used in this study undergo preliminary processing in advance to produce a flat-fielded image.  This processing includes statistical error initialization, conversion to count rates, subtraction of a dark rate image, division by a flat-field, and bad/hot pixel processing.

\begin{figure}
\plotone{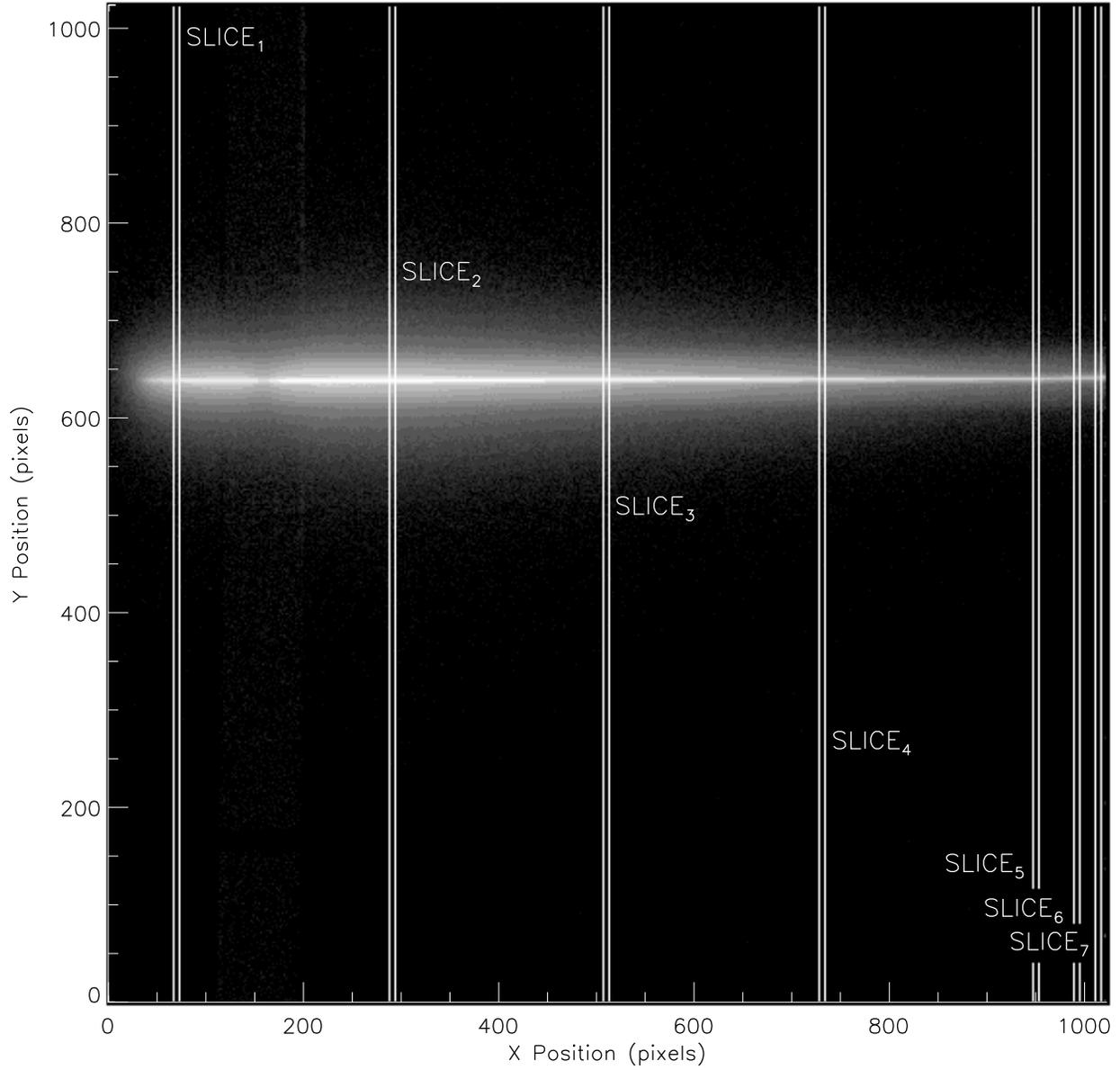}
\caption{Seven vertical crosscuts, $SLICE_{i}(y)$, are shown on the STIS/G140L spectral image of GD153. The 7-pixel wide by 1024-pixel long slices, centered at x$_{i}$=[69, 290, 511, 732, 953, 995, 1017], are summed in x.  \label{fig4}}
\end{figure}

\begin{figure}
\plotone{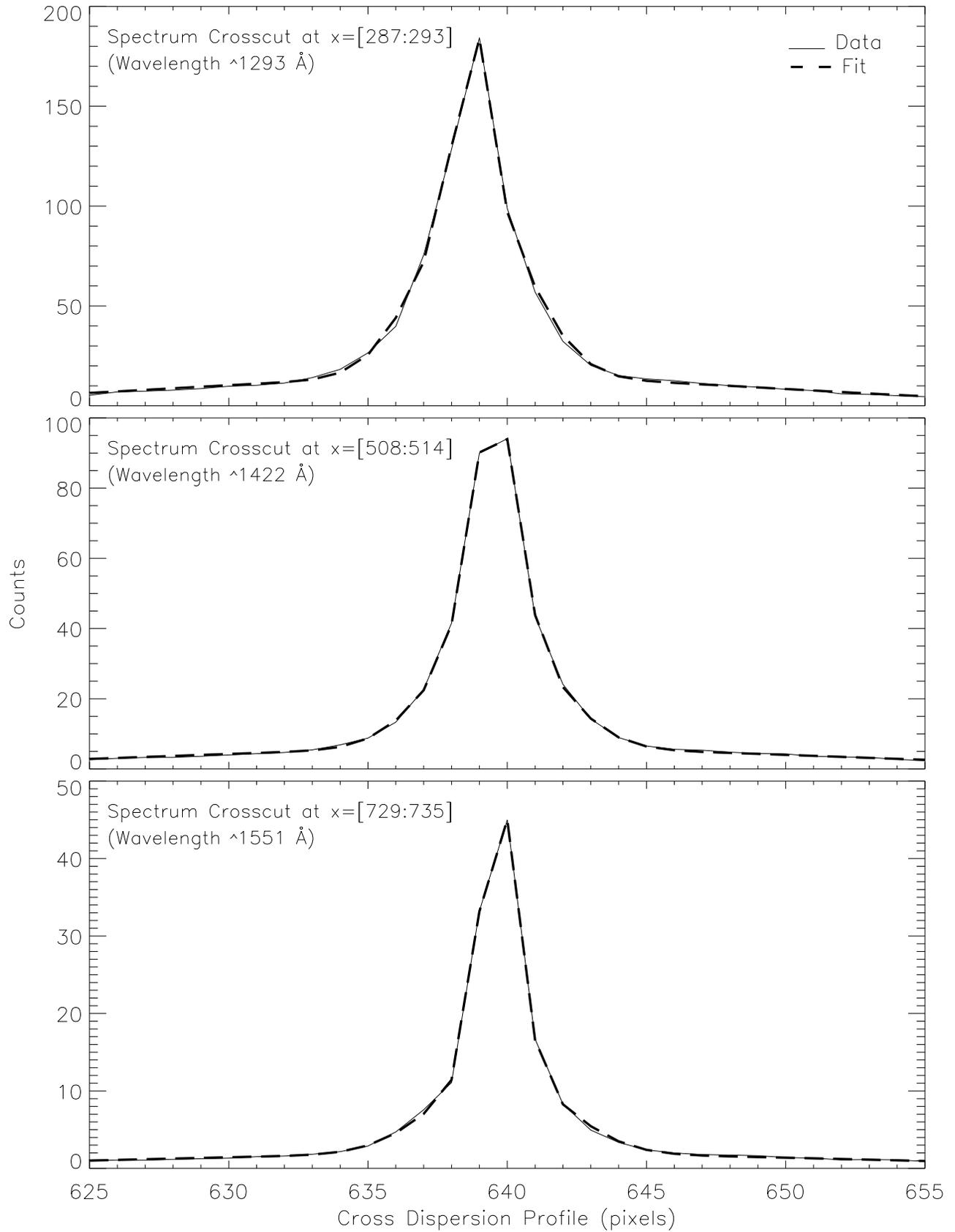}
\caption{Cross dispersion profiles of GD153 overplotted with our triple-Gaussian fits centered at three different x-locations.  The STIS data are marked with the solid lines.  Our triple-Gaussian functions are marked with the heavy, dotted lines.  \label{fig5}}
\end{figure}

Seven vertical crosscuts, $SLICE_{i}(y)$, are taken of the white dwarf spectrum in the flat-fielded, background-subtracted STIS image.  The subscript $i$ denotes the index (1-7) corresponding to a specific x-value, and the $y$ represents a 1024 element array perpendicular to the dispersion.  These seven crosscuts are each the sum of seven pixel wide slices centered at x$_{i}$=[69, 290, 511, 732, 953, 995, 1017].  These x-positions correspond to default, unshifted wavelengths of $\lambda_{i}$=[1164~\AA, 1293~\AA, 1422~\AA, 1551~\AA, 1680~\AA, 1705~\AA, 1718~\AA].  The spacing between these vertical slices at longer wavelengths is smaller because the response curve of the MAMA detector changes more rapidly at higher x-values (longer wavelengths).  (See Figure \ref{fig4}.)  Each of these slices is fit very accurately with a function represented by the sum of three Gaussians, $YFIT_{i}(y)$, each centered on the peak of the spectrum.  See Figure \ref{fig5} for three of the CDP crosscuts ($SLICE_{2},~SLICE_{3},~SLICE_{4}$) overplotted with heavy dotted curves denoting our triple-Gaussian fits ($YFIT_{2},~YFIT_{3},~YFIT_{4}$).  Each $YFIT_{i}(y)$ is then normalized by dividing by the total integrated area of its triple-Gaussian profile.  These seven normalized fits, $NORM_{i}(y)$, determine the shape of the stellar spectrum perpendicular to the dispersion (CDP) as a function of x (wavelength).  These normalized triple-Gaussian profiles represent our new weighted extraction slit and replace the box-car slit used in standard STIS data processing.

\subsection{DEVELOPING OUR DATA REDUCTION PROCEDURE}

%Once again, a simplified flowchart delineating the data reduction procedure development is provided.  See Figure \ref{fig6}. 

\begin{figure}
%\epsscale{1.0}
\plotone{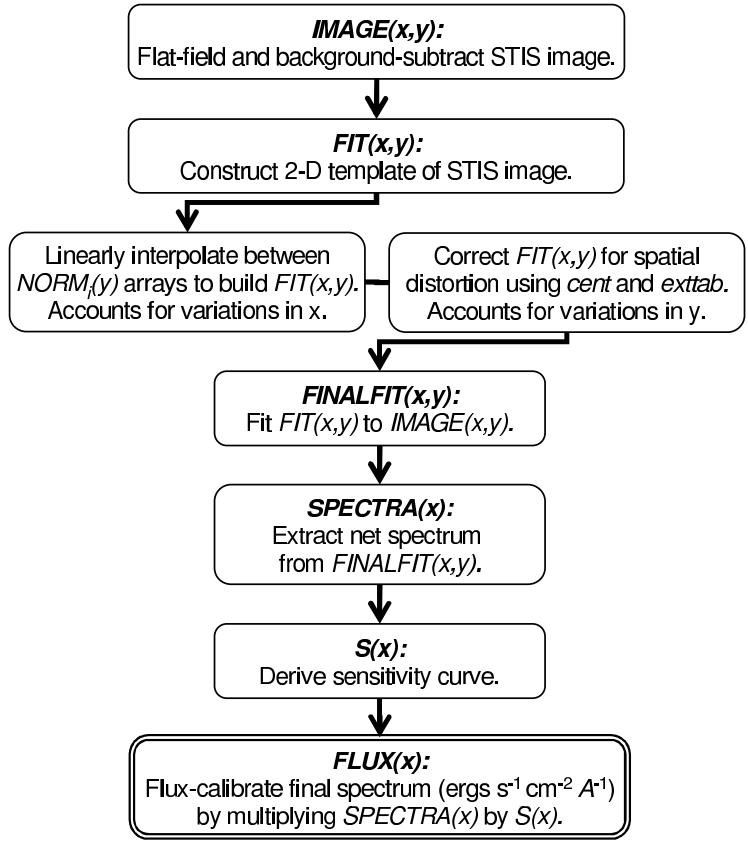}
\caption{Developing our data reduction procedure for a STIS/G140L spectral image utilizing the GD153 dataset.  (See Section 3.2 for further detail.) \label{fig6}}
\end{figure}

In this section, we describe the development of our data reduction procedure, STISFIT, written by C. L. Miskey in the Interactive Data Language (IDL).  We utilize this set of programs for the subsequent spectral extractions.  The simple flowchart in Figure \ref{fig6} outlines the development process for which we provide a more detailed discussion below.

We use our software first to perform the background subtraction for the region in the STIS image surrounding the stellar spectrum of GD153, which is trivial for data with a single star in the aperture.  Two background slices (one above and one below), each 5 pixels wide by 1024 pixels long, are taken from areas immediately adjacent to the spectrum, but far enough removed not to be contaminated by stellar flux.  This background array is averaged and then subtracted from the data.

Second, a generic template of a STIS spectral image is constructed using the $NORM_{i}(y)$ arrays.  A two-dimensional template array is built, $FIT(x,y)$, 1024 pixel columns by 1024 pixel rows.  Each $i$ of the $NORM_{i}(y)$ is associated with a proper wavelength.  Since GD153 is centered in the aperture, we use the default, unshifted, heliocentric wavelength scale, $W(x)$, spanning 1150-1736~\AA.  Each x (or wavelength) column of the template is individually created by linearly interpolating between corresponding values in the two bracketing vertical fits, $NORM_{i-1}(y)$ and $NORM_{i}(y)$.  For values of x$\le$69 ($i$=1), $NORM_{1}(y)$ is used.  Likewise, $FIT(x\ge1017,y)=NORM_{7}(y)$.

The previous step addresses variations in the CDP across the array in x. However, we must also consider corrections in the y-direction.  

\begin{figure}
\plotone{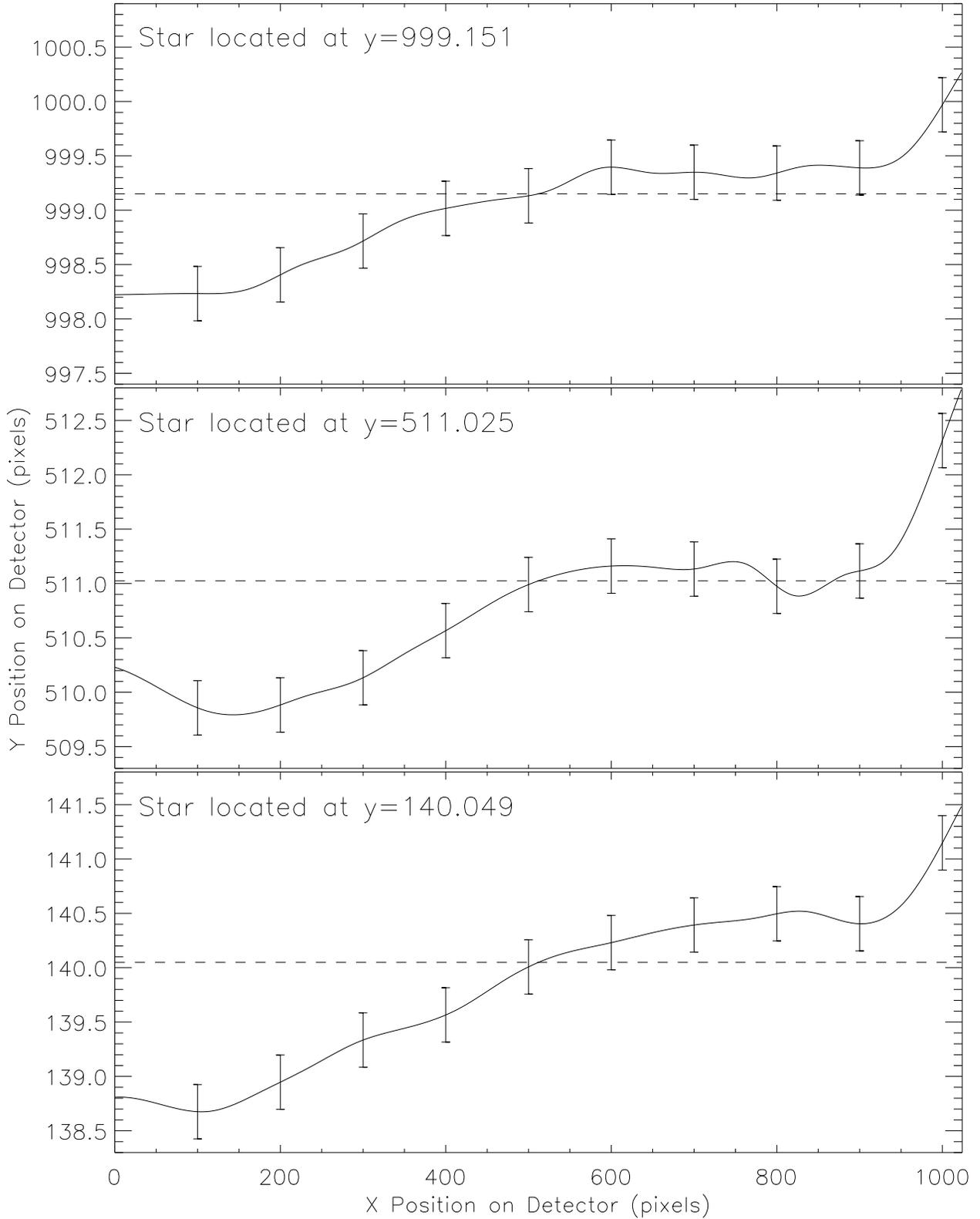}
\caption{The spatial distortion of a spectrum across the 1024 $\times$ 1024 pixel MAMA detector using the G140L grating for three different y-positions.  For approximate wavelength correlations, the central wavelength of the grating is 1425~\AA\ (x=512), and the average dispersion is 0.6~\AA/pixel.  \label{fig7}}
\end{figure}

A STIS stellar spectrum is not straight, but has a curvature that varies along the dispersion.  This curvature is not fixed, but also varies relative to y-position on the detector.  This spatial distortion has been accurately delineated by Don Lindler, who has compiled a reference file, {\it exttab}, for the G140L grating containing distortion curves for 27 different positions along the y-axis \citep[]{8}.  Similar {\it exttab} files assembled by Lindler exist for all of the other STIS gratings.  The curvatures for y-positions intermediate to those positions in the table are calculated through linear interpolation.  We define the y-position of a spectrum, represented by the variable $cent$ in our software, as the y-location of the center of the spectrum at x=512 in the 1024 $\times$ 1024 pixel STIS spectral image.  This definition is used throughout the paper.  Using the G140L {\it exttab} file and the y-location of GD153 on the detector, the template array, $FIT(x,y)$, can easily be corrected for spatial distortion.  The location of the spectral peak in $FIT(x,y)$ is mapped to the calculated distortion for each column of the array.  The spatial distortion across the STIS detector for three different y-positions is plotted in Figure \ref{fig7}.  A horizontal, dotted line runs through the star's central location at x=512.  Note that the deviations in y of the spectrum are most pronounced at extreme low and high values of x. 

Finally, this corrected data template is applied to the STIS image for GD153.  Each of the 1024 columns of the $FIT(x,y)$ array is fit to each of the 1024 columns of the flat-fielded, background-subtracted data, $IMAGE(x,y)$, using a linearization of the fitting function, to produce an array containing the model spectrum, $FINALFIT(x,y)$.  The routine generating $FINALFIT(x,y)$ uses the same technique as Bevington's CURFIT \citep[]{1}.  Using the notation of Bevington, for a function Y (the column of $FIT(x,y)$ being fit) of parameters A(I), the routine uses a linear expansion of Y as a function of A(I).  The array A has $n$ values (where $n$ equals the number of stellar sources; here $n$=1) denoting the amplitude(s) of the triple-Gaussian function(s).  The Y is evaluated for a set of parameters A and the array of independent variables X.  The procedure uses generalized least-squares to find the parameter increments which minimize $\chi^2$.  The program automatically iterates through parameter step sizes, which are controlled by input parameters.  Standard deviation errors are approximated by the inverse of the curvature matrix, (1/$\sigma^2$).  

The resulting fits show extremely good correlation.  We calculate the least squares fit between the actual, background-subtracted image and our STISFIT model image.  \\
\\
$Least~squares =  \displaystyle{\frac{\sum_{x=1}^{1024}{\sum_{y=-50}^{50}{[IMAGE(x,cent+y)-FINALFIT(x,cent+y)]^2}}}{\sum_{x=1}^{1024}{\sum_{y=-50}^{50}{[IMAGE(x, cent+y)]^2}}} \times 100\%}$. \\
\\
For GD153, a result of 0.39\% is obtained.  

The STISFIT net spectrum, $SPECTRA(x)$, is then created by summing an eleven pixel wide horizontal slice of the $FINALFIT(x,y)$, centered on the stellar y-location.  This extraction is not a straight horizontal cut, but follows the curvature on the detector characterized by the {\it exttab} file, as described above.  $SPECTRA(x)$ is in units of counts~second$^{-1}$~pixel$^{-1}$. 

We derive a new sensitivity curve, $S(x)$, by requiring that our spectral extraction yield the same measured fluxes versus wavelength for GD153 as the CALSTIS extraction.  CALSTIS is the standard STIS processing routine used in the Laboratory for Astronomy and Solar Physics at NASA/Goddard Space Flight Center \citep[]{8}.  Also, a very similar version of CALSTIS is used in the STScI pipeline processing \citep[]{3}.  Our STISFIT sensitivity curve, $S(x)$, is determined by dividing the GD153, flux-calibrated spectrum generated by CALSTIS by our net STISFIT spectrum, $SPECTRA(x)$.  This curve is dependent on wavelength.

With x corresponding to the unshifted wavelength array, $W(x)$, the STISFIT sensitivity curve is multiplied by the STISFIT generated net spectrum to yield the final, flux-calibrated, stellar spectrum; $S(x) \times SPECTRA(x) = FLUX(x)$.  The flux spectrum is in units of ergs~sec$^{-1}$~cm$^{-2}$~\AA$^{-1}$.  

Upon inspection of Figure \ref{fig8}, one sees that our final flux spectrum of GD153 differs slightly from the CALSTIS spectrum at low wavelengths.  This is because the MAMA detector has low sensitivity in this region, which leads to low signal-to-noise for the data and to a correspondingly less accurate sensitivity curve in the region shortward of 1200~\AA.  Therefore, we have chosen to smooth the derived $S(x)$ through this region to reduce the effects of noise and produce a more reliable sensitivity curve at these wavelengths.

\subsection{ASSESSMENT OF RESULTANT SPECTRA}

To test how closely our STISFIT procedure reproduces CALSTIS spectra, we calculate the percentage difference and average deviation between the extracted flux spectra generated by these two methods.  Here, the percentage difference and average deviation are given by:  \\
\\
$Percentage~difference = \displaystyle{\frac{\sum_{x=1}^{1024}{[CALSTIS(x)-FLUX(x)]}}{\sum_{x=1}^{1024}{CALSTIS(x)}} \times 100\%}$   \\
and
\\
$Average~deviation = \displaystyle{\frac{\sum_{x=1}^{1024}\sqrt{{[CALSTIS(x)-FLUX(x)]^2}}}{\sum_{x=1}^{1024}{CALSTIS(x)}} \times 100\%}$.  \\
\\
We use the STIS/G140L datasets of three stellar objects,  GD153, NGC~346-368, and Sand~3.  All are stellar point sources, but each has a different type of spectrum.  As previously stated, GD153 is a hot, white dwarf, continuum source with few photospheric or interstellar lines.  The object NGC~346-368 is a luminous O star in the Small Magellanic Cloud with a rich photospheric absorption spectrum, while Sand~3 is a hot, pre-white dwarf with a Wolf-Rayet emission line spectrum.  (See Figure \ref{fig8}.)  For the percentage differences, we obtain -0.06\%, 2.34\%, and 1.20\%, respectively.  The percentage difference indicates how closely the total flux is reproduced between the CALSTIS and STISFIT spectra.  The small values indicate that the total flux is conserved to within 2.5\% in our worst case.  These differences are in accord with the variations in sensitivity of the MAMA/G140L \citep[]{9}\footnotemark.  \footnotetext{STIS Instrument Science Reports are available at:  http://www.stsci.edu/hst/stis/documents/isrs.}  We expect a very small value for GD153, since it is the spectrum used to construct the STISFIT procedure, and one of our requirements of $S(x)$ is that STISFIT yield the same flux values as CALSTIS.  The insignificant difference is due to smoothing of the STISFIT sensitivity curve at lower wavelengths ($\lambda <$ 1200~\AA).  The largest variation is for an object with a rich line spectrum.  For the average deviation, we calculate 0.64\%, 3.77\%, and 2.82\%, respectively.  These results give us sufficient confidence in our procedures to apply them to our NGC~604 dataset.

\subsection{APPLYING OUR DEVELOPED MODELS AND SOFTWARE TO THE STIS DATASET FOR NGC~604}

%A simplified flowchart is again supplied.  See Figure \ref{fig9}, which outlines the application of our procedure to the NGC~604 dataset. 
%In this section, we describe the application of our data reduction procedure to our NGC~604 dataset. Once again, we include a simple flowchart in Figure \ref{fig9}, which outlines the process discussed in more detail below.

We apply the newly developed routines, STISFIT, to our STIS/G140L spectral image of NGC~604.  This set of programs has the capability to produce extracted, flux and wavelength calibrated, spectra for clusters of stars as well as isolated stars.  The final calibrated spectra are produced in a two step process.  We include a simple flowchart in Figure \ref{fig9}, which outlines the reduction procedure described in detail below.

%We describe the reduction procedure in detail below.

The first step is to use STISFIT to create an extracted, uncalibrated spectrum for each star.  We start by removing the background from the flat-fielded STIS image.  The sky background has several major contributions.  The geocoronal emission is comprised of the diffuse H~I-Ly$\alpha$ at 1216~\AA\ and the much weaker O~I~$\lambda$1304.  These geocoronal emission lines fill the entire 2$\arcsec$ wide slit in the spatial dimension and yield two vertical bands, one bright and the other much fainter, in the spectral image.  The H~I-Ly$\alpha$ band is quite distinct in the spectral image displayed in Figure \ref{fig2}.  The weak O~I emission is at most 10\% of the H~I-Ly$\alpha$ in brightness, depending on the time of the observation and the position of the target relative to the Earth.  In our observation, O~I is negligible.  

Because the scattered light from dust is essentially scattered UV stellar flux, its spectral characteristics mimic the spectra of the adjacent stars.  Consequently, care must be taken in applying any background subtraction.  This is especially the case for stars with low UV flux, where the accuracy of the background subtraction has a larger effect on the net spectrum.  In the UV wavelength range sampled by the G140L, nebular emission lines appear to be a negligible contribution to the background.

The most variable procedure in our data reduction scheme is the background subtraction.  Ideally, one wants to determine the background as close to the stellar source as possible.  Depending upon the degree of crowding in the FOV, different techniques may need to be applied.  As long as there is a reasonable amount of area free of stellar flux in the image, the standard background subtraction, averaging two bordering background strips, can be performed.  However, for a spectral image that contains very little area uncontaminated by stellar flux, a background image must be tailor-made.  Our NGC~604 image presents such a severe case.

In our NGC~604 spectral image, the densely populated stellar field makes it impossible to obtain direct background measurements adjacent to most of the individual spectra and subtract the background in the usual manner.  Hence, we create a 1024 $\times$ 1024 pixel array of the underlying background to be subtracted.  To map the background relative to wavelength, two horizontal strips of the STIS image (one in the lower portion and one in the upper portion of the dataset), 1024 pixels long by 10 pixels wide, are extracted and averaged from regions of the image containing no detectable UV stellar sources.  The section containing the geocoronal H~I-Ly$\alpha$ emission is assigned zero weight, and a 6th order polynomial is fit along x to the averaged background strip.  The H~I-Ly$\alpha$ band is then modeled with 2nd order polynomials for the regions corresponding to the edges of the 2$\arcsec$ wide aperture and a mean constant value denoting a flat plateau for the top.  These curves are combined to form the model, continuous, functional fit of the background in the x-direction, $BX(x)$.  Several vertical crosscuts of the image are then taken, and the minimum background values in the y-direction are measured to create an array of background levels perpendicular to the dispersion, $BY(y)$.  The two arrays $BX(x)$ and $BY(y)$ are combined, yielding a complete two-dimensional model, $B(x,y)$, for the background of the STIS spectral image of NGC~604.  We find this method of background determination quite reliable.

After the background model is subtracted from the image, STISFIT begins the spectral extractions.  It is very important to accurately determine the location of the center of each stellar spectrum, $cent(n)$, preferably on the sub-pixel level.  This is necessary to properly apply our triple-Gaussian spectral extraction algorithm.  Isolated spectra are extracted individually for faster processing.  For regions where a group of stellar spectra overlap or contaminate the local background, these spectra are all modeled together.  The variable $n$ equals the number of stars in each extraction set.  For each grouping, a $FIT(x,y,n)$ array is assembled with $n$ triple-Gaussian functions representing each of the point sources, corrected for stellar locations in the aperture and spatial distortions on the detector face.  Next, the template $FIT(x,y,n)$ is fit to the flat-fielded, background-subtracted $IMAGE(x,y)$.  A model for each spectrum, $SPECTRA(x,n)$, is then extracted from the grouped $FINALFIT(x,y,n)$.

The second step is to calculate the wavelength offset of each star and create the final, extracted, wavelength and flux calibrated, spectra.  First, strong interstellar lines are identified from the uncalibrated spectra.  These sharp, identified, spectral features serve as fiducial reference wavelengths.  They are used to calculate $shft(n)$, the x-position (wavelength) offset relative to the default, where the star is centered in the aperture.  Our corrected, heliocentric, wavelength array, $W_{\lambda}(x,n)$, is determined by using $shft(n)$ to shift the initial wavelength array, $W(x)$, for each spectrum accordingly.  The $S(x)$ array is mapped onto the new wavelength scales, producing the corrected sensitivity curves, $S_{\lambda}(x,n)$.  The $\lambda$ subscript denotes that the x array correlates to a shifted wavelength scale.  

The $shft(n)$ array is then used in a second run of our fitting routine, STISFIT, to  properly align the $NORM_{i}(y)$ arrays with their corresponding shifted wavelengths when the initial $FIT(x,y,n)$ template is built.  The shifted $FIT(x,y,n)$ is fit to $IMAGE(x,y)$, yielding $FINALFIT(x,y,n)$.  The new $SPECTRA(x,n)$ is extracted.  The shifted sensitivity curves, $S_{\lambda}(x,n)$, are multiplied by the results from the second pass of STISFIT, $SPECTRA(x,n)$, yielding the final, wavelength (\AA) and flux (ergs~sec$^{-1}$~cm$^{-2}$~\AA$^{-1}$) calibrated, stellar spectra, $FLUX(x,n)$. 

\section{COMPARISONS OF OUR DEVELOPED PROCEDURE WITH STANDARD STIS PROCESSING}

\subsection{THE SPECTRAL RESOLUTION OF EXTRACTED SPECTRA}

\begin{figure}
\plotone{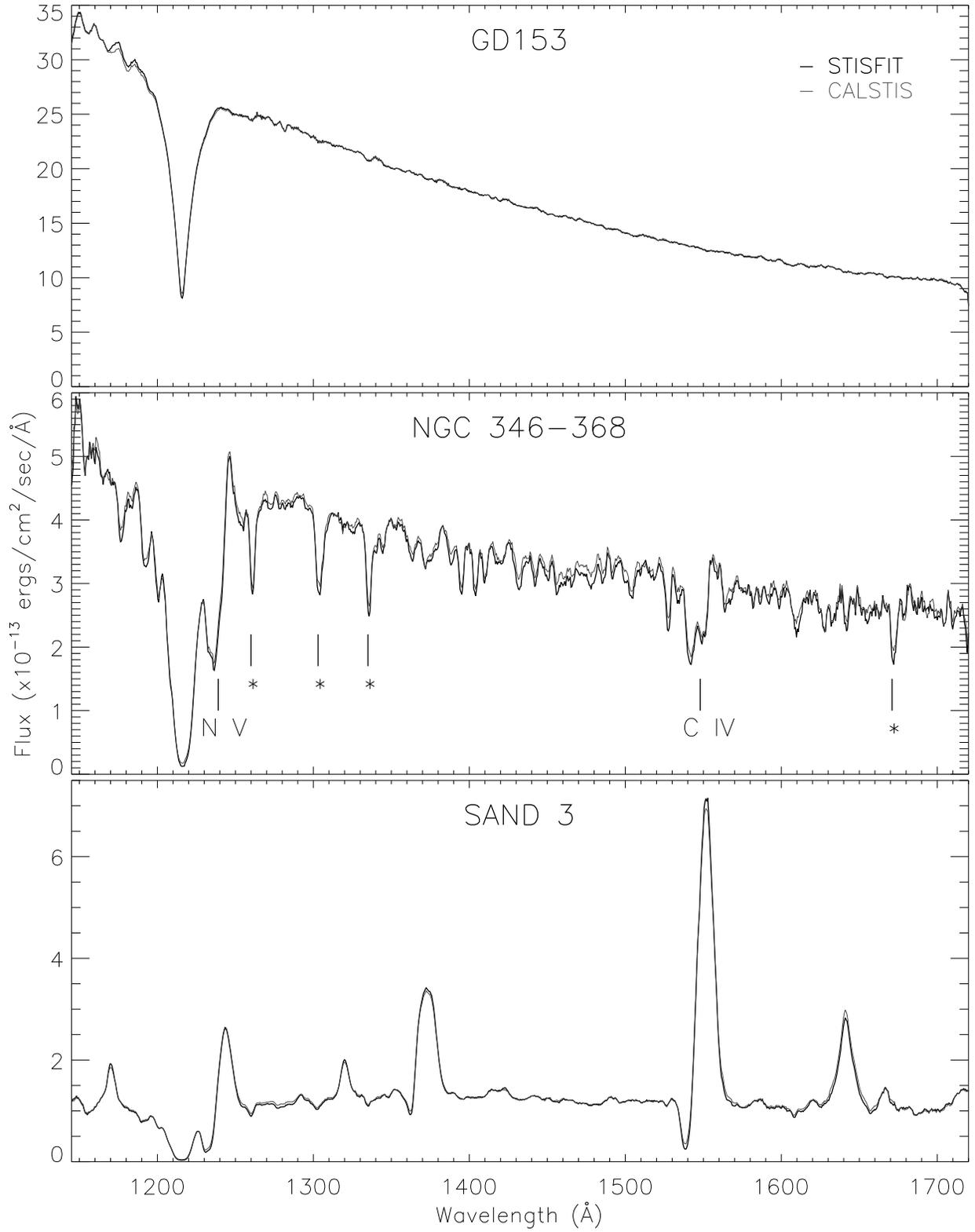}
\caption{Spectral extractions utilizing our STISFIT and the standard CALSTIS routines for three stellar objects.  N~V, C~IV, and several interstellar lines, marked with a *, are labeled in the middle panel.  The STISFIT data are indicated by the heavy, black, solid lines, while the CALSTIS data are shown with lighter lines.  \label{fig8}}
\end{figure}

\begin{figure}
%\epsscale{1.0}
\plotone{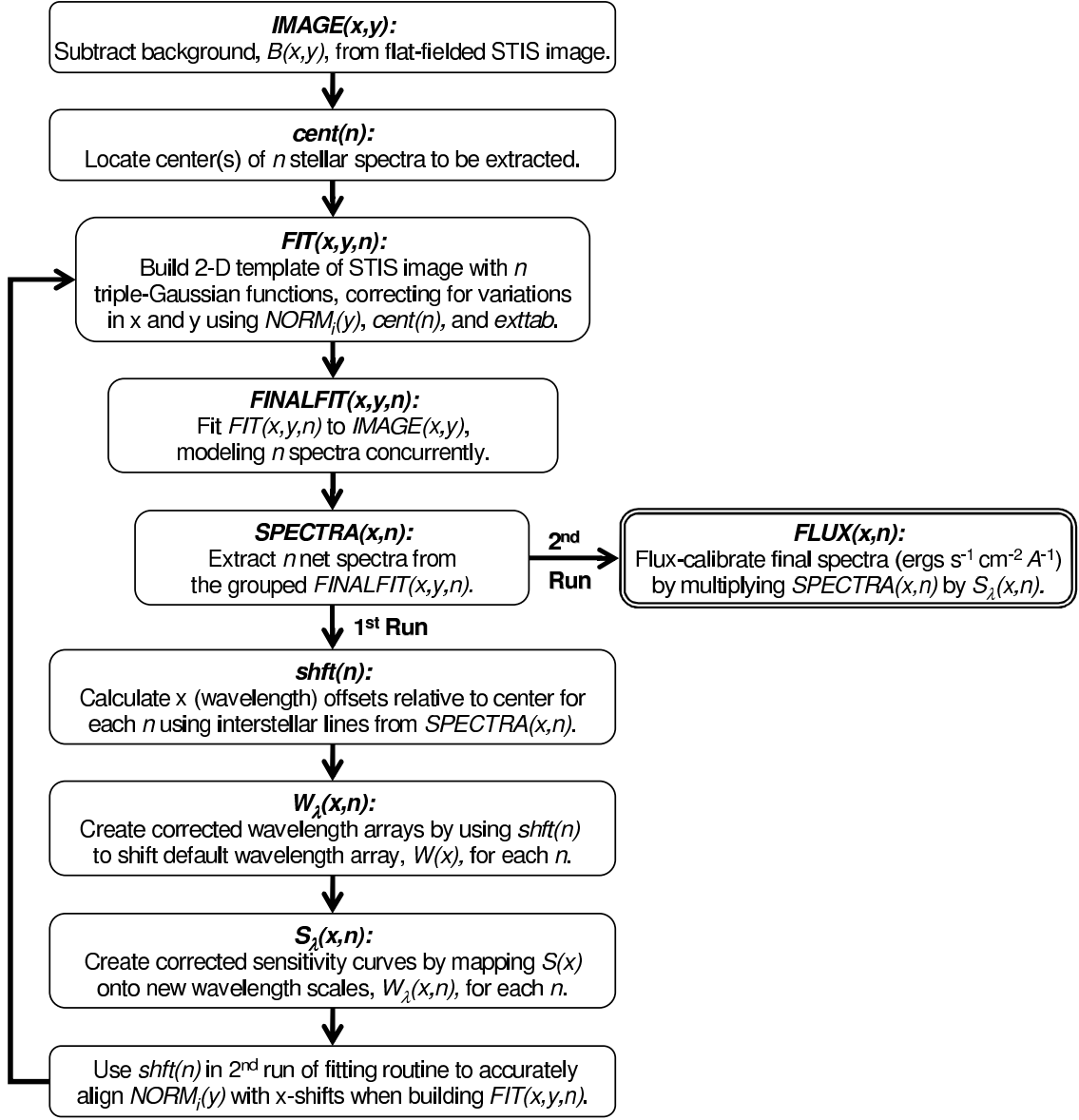}
\caption{Applying our developed models and software to the STIS dataset for NGC~604.  (See Section 3.4 for further detail.) \label{fig9}}
\end{figure}

Upon close examination of Figure \ref{fig8}, one finds that our STISFIT extraction, utilizing the triple-Gaussian fit to the cross dispersion profile as represented in Figure \ref{fig5}, produces deeper, sharper, absorption features than the CALSTIS extraction.  These differences are most apparent in the comparison of the processing of the G140L data for the bright O star, NGC346-368, in the H II region, NGC~346, of the SMC.  This is displayed in the middle panel of Figure \ref{fig8}.  Likewise, narrower, sharper N~V~$\lambda$1240, Si~IV~$\lambda$1400, C~IV~$\lambda$1550, and He~II~$\lambda$1640 profiles are also seen in the STISFIT spectrum in the comparison for the strong emission line object, Sand~3, in the bottom panel of Figure \ref{fig8}.

The slightly sharper features in our STISFIT extractions imply that STISFIT yields improved spectral resolution over the CALSTIS processing.  This might be expected, since our triple-Gaussian function gives higher weight to data near the center of the spectrum, whereas the box-car extraction used in CALSTIS gives equal weight to all datapoints.  The CDP can be thought of as the total contribution of the individual point spread functions (PSFs) produced at every wavelength along the spectrum.  When using a narrow slit, the PSF contribution is limited to the truncated portion that passes through the slit and is imaged on the detector.  This restricts the effects of the PSFs at greater distances in the dispersion direction.  The contribution from non-local PSFs at any point on the CDP away from the spectral peak, perpendicular to the dispersion, drops off dramatically due to truncation from the narrow slit.  However, the spectral images discussed here were obtained with the 2$\arcsec$ wide slit.  The PSFs are not truncated, unlike the case for the narrow slit.  Thus, at any specific wavelength, $\lambda_{\rm{o}}$, as one moves away from the center of the spectrum, perpendicular to the dispersion, there are higher fractional contributions to the flux from wavelengths farther from $\lambda_{\rm{o}}$.  Since our triple-Gaussian extraction slit gives higher weight to points near the center of the spectrum, the derived STISFIT fluxes at $\lambda_{\rm{o}}$ are determined by points much more local to $\lambda_{\rm {o}}$ than for the box-car extraction fluxes.  Other similar diffraction effects are discussed in Chapter 13 of the STIS Instrument Handbook \citep[]{7} and in  \citet[]{2}.

To quantitatively determine if STISFIT produces higher resolution data, one can compare both low resolution G140L STISFIT and CALSTIS extractions with a higher resolution E140M CALSTIS extraction for the same object.  The STIS/E140M echelle grating (R $\sim$ 10,000) spans the same wavelength range as the G140L.  The better G140L extraction should yield spectral line measurements more consistent with those obtained from the E140M spectrum.  For our comparison, we select the star NGC~346-368, since it has been observed with both gratings.  We compare the three extractions, and, as expected, the E140M data clearly show much sharper spectral features than those of both G140L extractions.  We measure the equivalent widths of the strong interstellar lines in all three extracted spectra.  For the G140L data, the STISFIT spectrum yields larger equivalent widths than the CALSTIS spectrum.  The STISFIT equivalent widths agree better with those measured in the E140M dataset, but are still systematically smaller.  This comparison indicates that the STISFIT extracted spectra provide better quantitative measurements for critical spectral parameters such as equivalent width. 

\begin{figure}
\plotone{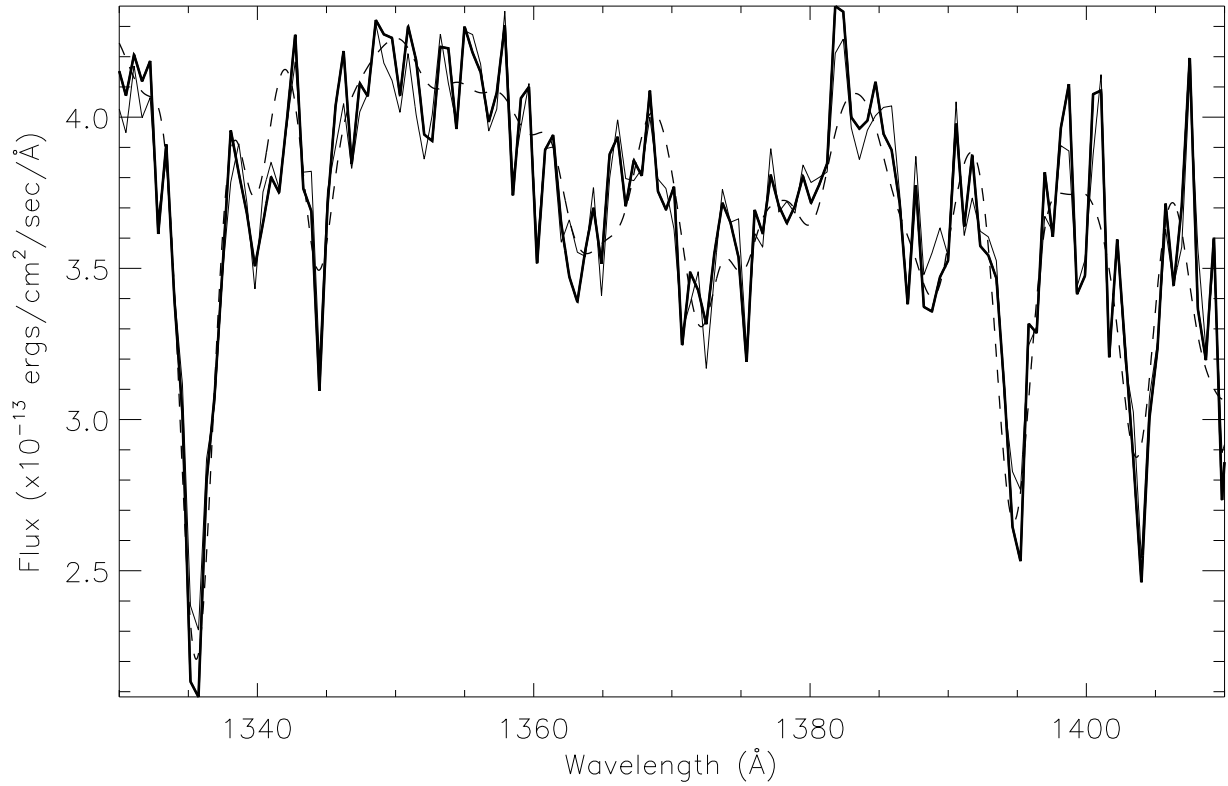}
\caption{Comparison between our STISFIT low-dispersion extraction, CALSTIS low-dispersion extraction, and convolved CALSTIS high-dispersion extraction for NGC~346-368.  The high-dispersion data are smoothed with a Gaussian to match the spectral resolution of the STISFIT data.  The STISFIT spectrum is plotted with the thick line. The CALSTIS low-dispersion spectrum is plotted with the thin line and the high-dispersion spectrum is plotted with the dotted line.    \label{fig10}}
\end{figure}

We also estimate the actual spectral resolution for the two G140L extractions. We do this by convolving (smoothing) the E140M data with a single Gaussian profile and correlating the result to the STISFIT and CALSTIS datasets.  The full width half maximum (FWHM) of the Gaussian smoothing is varied to obtain a least-square minimum (minimizing $\Sigma\sigma^2$).  The best matches to the STISFIT and the CALSTIS spectra are achieved if the E140M spectrum is convolved with Gaussians with FWHMs equal to 2.2~\AA\ (R $\sim$ 640) and 3.0~\AA\ (R $\sim$ 470), respectively.  (See Figure \ref{fig10}.)  Since the unsmoothed, intrinsic resolution for the E140M grating is a factor of 15 to 21 times higher, these two deduced values are good estimates of the spectral resolution near 1380~\AA.  Once again, our analysis suggests a higher resolution for the STISFIT extraction.

All resolution tests were performed on datasets obtained with the 2$\arcsec$ wide aperture.  We have not used our developed software to reduce any narrow slit STIS data.  Hence, we do not know if, given the effects of PSF truncation, our STISFIT technique would still yield a higher resolution in such cases.

\subsection{CLOSELY SPACED \& OVERLAPPING SPECTRA}

Perhaps the most important feature of our STISFIT process is its ability to extract several closely spaced spectra concurrently.  By modeling each stellar spectrum with its own triple-Gaussian fit, we are able to successfully separate out each star's flux contribution and minimize cross-contamination between neighboring spectra.  The standard STIS processing method can only extract one spectrum at a time with an eleven pixel wide box-car extraction slit.  Hence, it has no means of dealing with multiple overlapping spectra.

\begin{figure}
\plotone{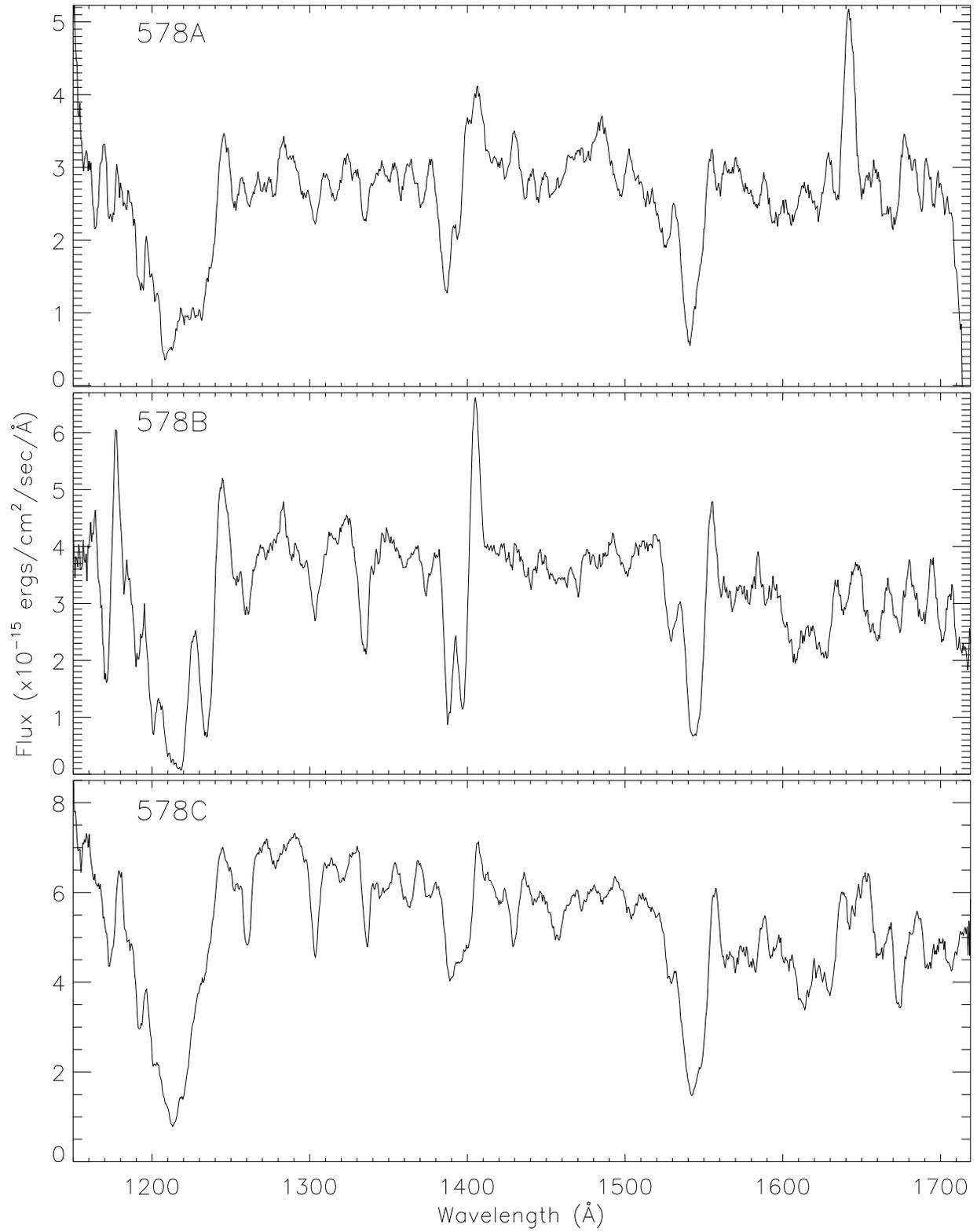}
\caption{STISFIT extracted spectra for the three closely spaced stars, 578A, 578B, and 578C, in NGC~604.  (Also see Figure \ref{fig2}.)  \label{fig11}}
\end{figure}

One excellent example in our NGC~604 dataset is the stellar complex near y=578, composed of three overlapping stars, 578A, 578B, and 578C.  (See Figure \ref{fig2}.)  Their spectral centers, $cent(n)$, are located at y=[578.700, 580.945, 583.626] in the image, corresponding to separations of 2.245 and 2.681 pixels, or 0.055$\arcsec$ and 0.065$\arcsec$, in the y-direction.  The STISFIT procedure produces three clean, low-noise spectra with very little cross-talk.  (See Figure \ref{fig11}.) Moreover, the spectral features and shape are consistent with what is expected for O9 stars.  This is even the case in the region beyond 1600~\AA, where the sensitivity of the MAMA falls off sharply. 

\section{DISCUSSION}

We have presented the basic techniques that compose our STISFIT extraction procedure for STIS spectral images of crowded stellar fields.  The foundation of our process is a triple-Gaussian extraction slit that gives higher weighting to flux values near the center of the spectrum.  As illustrated in Figure \ref{fig5}, the triple-Gaussian profile very closely fits the CDP, which enables us to determine the amount of flux corresponding to each stellar source for a group of overlapping spectra.  We have made use of the G140L spatial distortion reference file such that our extraction procedure can accurately track a spectrum at any position in the two-dimensional STIS spectral image.  The resulting STISFIT spectra yield deeper stellar absorption features.  Our tests indicate that the spectral resolution for STISFIT spectra are slightly better than that of the standard CALSTIS or STScI pipeline processing.  Furthermore, the strengths of the STISFIT extraction technique have allowed us to produce reliable, flux and wavelength calibrated, G140L spectra for stars separated by as little as 0.055$\arcsec$ in the direction perpendicular to dispersion.  This is at or near the spatial resolution limit of STIS.  Additional examples of our extracted spectra can be found in Paper 2.

We have recently expanded the STISFIT software to include the other low resolution grating configurations of STIS.  Our software routines are now applicable to spectral images obtained by both MAMA and CCD detectors, several different gratings, and the 52$\arcsec$ $\times$ 2$\arcsec$ and slitless aperture modes.  These improvements allow us to now obtain continuous spectral coverage, from ultraviolet through optical wavelengths, of individual stars in crowded fields.  We will present observational results and assessments of these techniques in following papers.

We wish to acknowledge the assistance and invaluable advice provided by Don Lindler, Nicholas Collins, Eliot Malamuth, and Charles Bowers; all are members of the STIS Instrument Development or Support Groups at NASA/Goddard Space Flight Center.  We also acknowledge support through NASA grant NAG5-3378.  The NGC~604 STIS spectral image was obtained under Guaranteed Observing Time awarded to Dr. Albert Boggess. 


\appendix
%\section{Data Reduction Flowcharts}
%\placefigure{fig9}
%\placefigure{fig10}
%\placefigure{fig11}

\newpage
\section{Glossary}

\begin{description}
%\begin{enumerate}
\item[\em{B(x,y)}] Two-dimensional model array of the background for our STIS spectral image.
\item[\em{BX(x)}] Model, continuous, functional fit of the background in the x-direction for our STIS dataset.
\item[\em{BY(y)}] Array of background levels perpendicular to the dispersion for our STIS dataset.
\item[\em{CALSTIS}] Standard STIS data reduction procedure used at LASP, NASA/GSFC and STScI.
\item[\em{CALSTIS(x)}] Flux-calibrated spectrum from a STIS image generated with CALSTIS.
\item[\em{CDP}] Cross dispersion profile.
\item[\em{cent(n)}] Y-location in pixels of the center of $n$ spectra at x=512 in a 1024$\times$1024 pixel STIS spectral image.
\item[\em{exttab}] Reference file containing spatial distortion curves for STIS gratings.
\item[\em{FINALFIT(x,y,(n))}] Resultant array containing the model spectra created by fitting our STIS template, $FIT(x,y,(n))$, to the spectral image, $IMAGE(x,y)$.
\item[\em{FIT(x,y,(n))}] Generic template of a STIS spectral image constructed with $NORM_{i}(y)$, $cent(n)$, and $exxtab$.  Accounts for variations in both x and y on the STIS detector.
\item[\em{FLUX(x,(n))}] Final, flux-calibrated, stellar spectra from a STIS image generated with our STISFIT procedure in units of ergs~sec$^{-1}$~cm$^{-2}$~\AA$^{-1}$.  
\item[\em{IMAGE(x,y)}] Flat-fielded, background-subtracted, STIS image.
\item[\em{NORM$_{i}$(y)}] Seven normalized $YFIT_{i}(y)$ arrays, each a triple-Gaussian function delineating the cross dispersion profile of the STIS detector at a specific value of x.  Represents our new weighted extraction slit.
\item[\em{S(x)}] Our default STISFIT sensitivity curve.
\item[\em{S$_{\lambda}$(x,n)}] Shifted STISFIT sensitivity curves, corrected for stellar offset relative to default in the aperture. 
\item[\em{shft(n)}] The x-position (wavelength) offset relative to default, where the star is centered in the aperture, for each $n$.
\item[\em{SLICE$_{i}$(y)}] Seven vertical crosscuts, perpendicular to the dispersion, extracted from the STIS image of GD153.  Each is the sum of seven columns and corresponds to a specific x (wavelength) value. 
\item[\em{SPECTRA(x,(n))}] Net spectra from a STIS image generated with our STISFIT procedure in units of counts~second$^{-1}$~pixel$^{-1}$.
\item[\em{STISFIT}] Our set of data reduction routines, written in IDL, which extract flux and wavelength calibrated spectra from a crowded STIS field-of-view.
\item[\em{W(x)}] Default, heliocentric, wavelength scale.  For STIS/G140L, it  spans 1150-1736~\AA.
\item[\em{W$_{\lambda}$(x,n)}] Shifted, heliocentric, wavelength array, corrected for stellar offset relative to default in the aperture. 
\item[\em{YFIT$_{i}$(y)}] Seven resultant arrays from the fit of triple-Gaussian functions (the sum of three Gaussians, each centered on the peak of the spectrum) to the $SLICE_{i}(y)$ arrays.
\end{description}

\clearpage

\clearpage

\clearpage

\clearpage

\clearpage

\clearpage

\clearpage

\clearpage

\clearpage

\clearpage

\clearpage

%\end{center}

%% Tables should be submitted one per page, so put a \clearpage before
%% Two options are available to the author for producing tables:  the
%% deluxetable environment provided by the AASTeX package or the LaTeX
%% table environment.  Use of deluxetable is preferred.
%%
%% Three table samples follow, two marked up in the deluxetable environment,
%% one marked up as a LaTeX table.
%%
%% In this first example, note that the \footnotesize command has been
%% used to shrink the table so it will fit on one page. Note also that
%% the \label command needs to be placed inside the \tablecaption.

\clearpage

\begin{deluxetable}{ccccc}
\small
\tablewidth{0pt}
\tablecaption{Hubble Space Telescope Datasets Used}
\tablehead{
\colhead{Object} & \colhead{Instrument} & \colhead{Filter/Grating} & \colhead{Central $\lambda$} & \colhead{Dataset Name}\\  
\colhead{}& \colhead{}& \colhead{}& \colhead{(\AA)} & \colhead{}}
\startdata
GD153   & STIS &	G140L & 1425 &		o4a502060 \\ 
%\tableline
NGC~346 & STIS &	G140L & 1425 &		o46j01010 \\*
\nodata	& STIS &	E140M & 1425 &		o4wr01030 \\ 
%\tableline
NGC~604 & STIS &	G140L & 1425 &		o4x101040 \\*
\nodata	& WFPC2 &	F336W & 3344 &		u2ab0207t \\*
\nodata	& WFPC2 &	F336W & 3344 &		u2ab0208t \\*
\nodata	& WFPC &	F439W & 4353 &		w0nn0101t \\*
\nodata	& WFPC &	F439W & 4353 &		w0nn0102t \\*
\nodata	& WFPC &	F439W & 4353 &		w0nn0103t \\ 
%\tableline
SAND~3  & STIS &	G140L & 1425 &		o4x101030 \\ 
\enddata
\end{deluxetable}


\begin{thebibliography}{}
\bibitem[Bevington(1969)]{1}Bevington, P. R. 1969, Data Reduction and Error Analysis For The Physical Sciences (New York: McGraw-Hill, Inc.)
\bibitem[Bowers \& Baum (1998)]{2} Bowers, C. \& Baum, S. 1998, STIS Instrument Science Report 98-24
\bibitem[Brown (2002)]{3}Brown, T., et al. 2002, HST STIS Data Handbook, Version 4.0, (Baltimore: STScI)
\bibitem[Bruhweiler, Miskey, \& Smith Neubig (2003)]{4}Bruhweiler, F. C., Miskey, C. L., \& Smith Neubig, M. 2003, AJ, submitted (Paper 2)
\bibitem[Drissen, Moffat, \& Shara (1993)]{5}Drissen, L., Moffat, A. F. J., \& Shara, M. M. 1993, AJ, 105, 1400

\bibitem[Helou et al.(1995)]{6}  Helou, G., Madore, B. F., Schmitz, M., Wu, X., Corwin, H. G., Jr., Lague, C., Bennett, J., \& Sun, H.  1995, The NASA/IPAC Extragalactic Database, Information \& On-Line Data in Astronomy, ed. D. Egret \& M. A. Albrecht.  Ap\&SS, 203, 95
\bibitem[Leitherer et al.(2001)]{7}Leitherer, C., et al. 2001, STIS Instrument Handbook, Version 5.1, (Baltimore: STScI)
\bibitem[Lindler(1999)]{8}Lindler, D. 1999, CALSTIS Reference Guide, Version 6.4, (Greenbelt, MD: GSFC)
\bibitem[Stys \& Walborn (2001)]{9} Stys, D. J., \& Walborn, N. R. 2001, STIS Instrument Science Report 2001-01R

\end{thebibliography}
\end{document}